\documentclass[11pt,preprint]{aastex}

\def\vec#1{\mbox{\boldmath $#1$}}

\begin{document}

\title{The Role of the Equation of State in Resistive Relativistic Magnetohydrodynamics}

\author{Yosuke Mizuno}
\affil{Institute of Astronomy, National Tsing-Hua University 
Hsinchu 30013, Taiwan, R.O.C.; mizuno@phys.nthu.edu.tw}

\shorttitle{The Effect of EoS in RRMHD}
\shortauthors{Mizuno}

\begin{abstract}
We have investigated the role of the equation of state in resistive relativistic magnetohydrodynamics using a newly developed resistive relativistic magnetohydrodynamic code. A number of numerical tests in one-dimension and multi-dimensions are carried out in order to check the robustness and accuracy of the new code. The code passes all the tests in situations involving both small and large uniform conductivities. Equations of state which closely approximate the single-component perfect relativistic gas are introduced. Results from selected numerical tests using different equations of state are compared.  The main conclusion is that the choice of the equation of state as well as the value of the electric conductivity can result in considerable dynamical differences in simulations involving shocks, instabilities, and magnetic reconnection. 
\end{abstract}
\keywords{magnetohydrodynamics (MHD) - methods: numerical - plasmas - relativistic processes}

\section{Introduction}

Magnetic fields play an important role in determining the evolution of the matter in many astrophysical objects. In highly conducting plasma, the magnetic field can be amplified by gas contraction or shear motion. Even when the magnetic field is weak initially, the magnetic field can grow rapidly and influence the gas dynamics of the system. This is particularly important for the high-energy astrophysical phenomena related to strongly magnetized relativistic plasmas associated with objects such as active galactic nuclei (AGNs) (e.g., Urry \& Pavovani 1995), relativistic jets (e.g., Mirabel \& Rodr\'{i}guez 1999; Blandford 2002), pulsar winds (e.g., Gaensler \& Slane 2006; Kirk et al. 2009), gamma-ray bursts (Zhang \& M\'{e}sz\'{a}ros 2004; Piran 2005; M\'{e}sz\'{a}ros 2006), and magnetars (e.g., Woods \& Thompson 2006; Mereghetti 2008). 

The ideal magnetohydrodynamic (MHD) approximation is a good description of the global properties and dynamics of such systems well into their nonlinear regimes.  In this limit the electrical resistivity $\eta=1/\sigma$ vanishes (infinite electrical conductivity). In this framework, many multidimensional ideal relativistic MHD (RMHD) codes have been developed to investigate relativistic astrophysical phenomena including fully non-linear regimes  (e.g., Komissarov 1999; Koide et al. 1999; Komissarov 2001; Koldoba et al. 2002; Del Zanna et al. 2003; Leismann et al. 2005; Gammie et al. 2003; De Villiers \& Hawley 2003; Anninos et al. 2005; Duez et al. 2005; Shibata \& Sekiguchi 2005; Ant\'{o}n et al. 2006; Mignone \& Bodo 2006; Mizuno et al. 2006; Neilson et al. 2006; Del Zanna et al. 2007; Giacomazzo \& Rezzolla 2007; Farris et al. 2008; Mignone et al. 2009; Beckwith \& Stone 2011; Inoue et al. 2011). The ideal MHD limit provides a convenient form for solving the equations of RMHD and is also an excellent approximation for many relativistic astrophysical phenomena. However, in extreme cases such as binary mergers (the merger of two neutron stars or of a neutron star with a black hole) (e.g. Shibata \& Taniguchi 2011; Faber \& Rasio 2012) or the central engine of long GRBs (collapsar) (e.g., MacFadyen \& Woosley 1999) the electrical conductivity can be small, and regions of high resistivity may appear. 

Quite often numerical simulations using ideal RMHD exhibit violent magnetic reconnection. The magnetic reconnection observed in ideal RMHD simulations is due to purely numerical resistivity, occurs as a result of truncation errors, and hence fully depends on details of the numerical scheme and resolution. Magnetic reconnection is one of the most important phenomena in astrophysics. It is highly dynamic, and it converts magnetic energy into fluid energy. The magnetic reconnection process has been invoked to explain flaring events (e.g., Lyutikov 2006; Giannios et al. 2009) and magnetic annihilation (Coroniti 1990; Lyubarsky \& Kirk 2001) in relativistic plasmas.  Therefore, numerical codes  solving the resistive RMHD (RRMHD) equations  and that allow control of magnetic reconnection according to a physical model of resistivity are highly desirable.     

Numerical simulation using the ideal RMHD equations is considerably easier than using the RRMHD equations because the equations become mixed hyperbolic with stiff relaxation terms. The pioneering work on resistive RMHD done by Komissarov (2007) solved the numerical flux by using the Harten-Lax-van Leer (HLL) approximate Riemann solver and  by using Strang's splitting technique for the stiff relaxation terms. More recently, Palenzuela et al. (2009) have proposed a numerical method that solves the stiff relaxation terms in the equations by an implicit-explicit (IMEX) Runge-Kutta method, and Dionysopoulou et al. (2012) have extended the work of Palenzuela et al. (2009) to 3D.  A different approach has been taken by Dumbser \& Zanotti (2009) who have applied the high order $P_{N}P_{M}$ scheme to solving the resistive RMHD equations, and also Takamoto \& Inoue (2011) who have used the method of characteristics to solve the Maxwell equations accurately. Even more recently, a 3+1 resistive general relativistic MHD (GRMHD) code using mean-field dynamo closure has been developed by Bucciantini \& Del Zanna (2012).

Plasma in the relativistic regime can have three major characteristics: the system has (1) relativistic fluid velocity (kinetic energy much greater than rest-mass energy), has (2) relativistic temperature (internal energy much greater than rest-mass energy), or has (3) relativistic Alfv\'{e}n speed (magnetic energy much greater than rest-mass energy). The second characteristic of relativistic temperature brings us to the issue of the equation of state (EoS) of the plasma. The EoS most commonly used in RMHD simulations is designed for plasmas with constant specific heat ratio (the so-called ideal EoS). However, this ideal EoS is valid only for plasmas with either ultra-relativistic temperature or non-relativistic temperature. The theory of relativistic perfect gases (Synge 1957) has shown that the specific heat ratio cannot be constant if consistency with kinetic theory is required. However, the exact EoS involves modified Bessel functions and is too complicated to be efficiently implemented in numerical codes. To get around this problem Mignone et al. (2005) introduced an approximate EoS given by a simple analytical formulation in the context of relativistic non-magnetized flows. This approximate EoS was applied in the context of relativistic MHD by Mignone \& McKinney (2007). A  different EoS approximation than that proposed by Mignone et al. (2005) has been proposed by Ryu et al. (2006).  Clearly a determination of the effects of a difference in the choice of the approximate EoS is important to further advances in RRMHD.

In this paper, we present the development of a new resistive RMHD simulation code including different realistic EoS approximations such as those proposed by Mignone et al. (2005) or by Ryu et al. (2006). This new RRMHD code is based on the ideal RMHD code RAISHIN (Mizuno et al. 2006; 2011) which uses a Godunov-type scheme to solve the conservation equations of ideal RMHD. In particular, we apply this new code to the role of the EoS in the resistive RMHD regime.  
We describe the basic equations of resistive RMHD in \S 2, three different equations of state are investigated  in \S 3, and the numerical methods are described in \S 4.  The various numerical tests in one-dimension  and multi-dimensions are presented in \S 5. In \S 6 we conclude.

\section{Basic Equations of Resistive Relativistic MHD}

We have considered $n^{\alpha}$ to be the time-like translational killing vector field in a flat (Minkowski) space-time, so $n^{\alpha}=(-1,0,0,0)$, where we use Greek letters that take values from 0 to 3 for the indices of 4D space-time tensors, while Roman letters take values from 1 to 3 for the indices of 3D spatial tensors. We use the speed of light $c=1$ and Lorentz-Heaviside notation for electromagnetic quantities, so that all $\sqrt 4 \pi$ factors disappear.  

The total energy-momentum tensor $T^{\alpha \beta}$ describing a perfect fluid coupled to an electromagnetic field is defined as
\begin{equation}
T^{\alpha \beta}= T^{\alpha \beta}_{fluid} + T^{\alpha \beta}_{EM}.
\end{equation}
The first term is due to matter:
\begin{equation}
T^{\alpha \beta}_{fluid}= \rho hu^{\alpha}u^{\beta}+pg^{\alpha \beta},
\end{equation}
where $u^{\alpha}$ is the fluid four-velocity, while $h$ $(=1+ \epsilon + p/\rho)$, $\rho$, $p$ and $\epsilon$ are the enthalpy, the proper rest mass density, the gas pressure, and the specific internal energy as measured in the fluid rest frame. The second term comes from the electromagnetic field:
\begin{equation}
T^{\alpha \beta}_{EM} = F^{\alpha \mu} F_{\mu}^{\beta} - {1 \over 4} (F^{\mu \nu}F_{\mu \nu})  g^{\alpha \beta},
\end{equation}
where $F^{\alpha \beta}$, and its dual $^*F^{\alpha \beta}$ are the Maxwell and Faraday tensors of the electromagnetic field given by
\begin{eqnarray}
&& F^{\alpha \beta} =  n^{\alpha} E^{\beta} - n^{\beta} E^{\alpha} + n_{\nu} e^{\nu \alpha \beta \mu} B_{\mu}, \\
&& ^*F^{\alpha \beta} =  n^{\alpha} B^{\beta} - n^{\beta} B^{\alpha} + n_{\nu} e^{\nu \alpha \beta \mu} E_{\mu}. 
\end{eqnarray}  
$E^{\alpha}$ and $B^{\alpha}$ are the electric and magnetic fields as measured by an observer moving along any time-like vector $n^{\alpha}$, while $e^{\alpha \beta \mu \nu} = \sqrt{-g} \epsilon_{\alpha \beta \mu \nu}$ is the Levi-Civita alternating tensor of space-time and $\epsilon_{\alpha \beta \mu \nu}$ is the four-dimensional Levi-Civita symbol. 

In the global inertial frame with time-independent coordinate grid, the full system of Euler and Maxwell's equations are
\begin{eqnarray}
&& \partial_{t} D + \nabla \cdot D\vec{v} =0, \\
&& \partial_{t} \vec{m} + \nabla \cdot \vec{\Pi} = 0, \\
&& \partial_{t} \tau + \nabla \cdot \vec{Y} =0, \\
&& \partial_{t} \vec{E} - \nabla \times \vec{B} + \nabla \Psi =  -\vec{j}, \\
&& \partial_{t} \vec{B} + \nabla \times \vec{E} + \nabla \Phi = \vec{0}, \\
&& \partial_{t} \Psi + \nabla \cdot \vec{E} = q - \kappa \Psi, \\
&& \partial_{t} \Phi + \nabla \cdot \vec{B} = -\kappa \Phi, \\
&& \partial_{t} q + \nabla \cdot \vec{j} = 0,
\end{eqnarray}  
where $\vec{j}$ is the spatial current vector, $q$ is the charge density, $\kappa$ is the damping rate parameter, and 
the conserved variables
\begin{eqnarray}
D &=& \rho \gamma, \\
\vec{m} &=& \rho h \gamma^{2} \vec{v} + \vec{E} \times \vec{B}, \\
\tau &=& \rho h \gamma^{2} - p + {1 \over 2}(E^{2} + B^{2})
\end{eqnarray} 
express the relativistic mass density, the momentum density, and the total energy density. Here, $\vec{v}$ is the velocity measured by an inertial observer and $\gamma \equiv 1/\sqrt{1-v^{2}}$ is the Lorentz factor. The energy flux density and the momentum flux density can then be given by
\begin{eqnarray}
\vec{Y} &=& \rho h \gamma^{2} \vec{v} + \vec{E} \times \vec{B}, \\
\vec{\Pi} &=&  - \vec{E}\vec{E} - \vec{B}\vec{B} + \rho h \gamma^{2} \vec{v}\vec{v} + \left[ {1 \over 2} (E^{2}+B^{2}) + p \right] \vec{g} .
\end{eqnarray}
An equation of state (EoS) is needed to close the system, we have adopted a variable EoS (e.g., Mignone et al. 2005; Mignone \& McKinney 2007; Ryu et al. 2006). The details of the variable EoS are explained in next section.

Eqs. (9)-(12) evolve the augmented Maxwell's equations which contain two additional fields $\Psi$ and $\Phi$ to control the system dynamics. In this approach, the two scalar fields $\Psi$ and $\Phi$ indicate deviations of the divergence of the electric and magnetic fields from the values prescribed by Maxwell's equations, propagate at the speed of light, and decay exponentially over a time-scale $\sim 1/\kappa$ when the damping rate parameter $\kappa > 0$. Following previous studies (Komissarov 2007; Palenzuela et al. 2009; Dumbser \& Zanotti 2011), we have adopted the so-called hyperbolic divergence-cleaning approach used in the context of ideal MHD (Dedner et al. 2002). 

The system of Eqs. (6)-(13) is closed by means of Ohm's law. Ohm's law for relativistic plasmas can be very complicated (e.g., Lichnerowicz 1967; Ardavan 1984; Blackman \& Field 1993; Gedalin 1996; Melatos \& Melrose 1996; Punsley 2001; Meier 2004).  In this paper, we consider only the simplest kind of relativistic Ohm's law that assumes an isotropic plasma resistivity (e.g., Komissarov 2007; Palenzuela et al. 2009; Zenitani et al. 2010; Takamoto \& Inoue 2011; Takahashi et al. 2011). In covariant form, the four-vector of the electric current is obtained from
\begin{equation}
I^{\alpha} = \sigma F^{\alpha \beta} u_{\beta} + q_{0} u^{\alpha}, 
\end{equation}
where $\sigma=1/\eta$ is the conductivity, $\eta$ is the resistivity, and $q_{0} = - I_{\alpha}u^{\alpha}$ is the electric charge density as measured in the fluid flame (Lichnerowicz 1967; Blackmas \& Field 1993). In a special relativistic inertial frame, we find  
\begin{equation} 
\vec{j} = \sigma \gamma [ \vec{E} + \vec{v} \times \vec{B} - (\vec{E} \cdot \vec{v})\vec{v}] + q \vec{v}. 
\end{equation} 
In the fluid rest frame, this equation becomes
\begin{equation}
\vec{j} = \sigma \vec{E}.
\end{equation}
The ideal MHD limit of Ohm's law is given by the limit of infinite conductivity ($\sigma \to \infty$). In this limit Eq. (20) reduces to
\begin{equation} 
\vec{E}+\vec{v} \times \vec{B} - (\vec{E} \cdot \vec{v}) \vec{v} = 0.
\end{equation}
Splitting this equation into components normal and parallel to the velocity vector give
\begin{eqnarray}
&& \partial_{t} \vec{E}_{\parallel} + \sigma \gamma [\vec{E}_{\parallel} - (\vec{E} \cdot \vec{v}) \vec{v}] =0, \\
&& \partial_{t} \vec{E}_{\perp} + \sigma \gamma [\vec{E}_{\perp} + \vec{v} \times \vec{B}] = 0.
\end{eqnarray} 
From these equations, one obtains the well-known ideal MHD condition
\begin{equation}
\vec{E} = - \vec{v} \times \vec{B}.
\end{equation}
In this limit the electric field is orthogonal to both magnetic and velocity fields.

\section{Equations of State}

An EoS relating thermodynamic quantities is required to close the system of eqs. (6)-(13). In general, an EoS is written as
\begin{equation}
h \equiv h(p, \rho),
\end{equation}
and general forms for the polytropic index $n$ and the sound speed $c_{s}$ are given by
\begin{equation} 
n=\rho {\partial h \over \partial \rho} -1, \  c_{s}^{2} = - {\rho \over nh} {\partial h \over \partial \rho}.
\end{equation}

The most commonly used EoS, a constant $\Gamma$-law (ideal) EoS, is given by
\begin{equation} 
h= 1+ {\Gamma \over \Gamma -1} \Theta,
\end{equation}
where $\Gamma$ is the constant specific heat ratio and $\Theta =p/\rho$ is the temperature. 
The sound speed is calculated from
\begin{equation}
c_{s}^{2} = {\Gamma \Theta \over h}.
\end{equation} 
The constant $\Gamma$-law EoS may be applied correctly to a plasma with non-relativistic temperature where $\Gamma=5/3$ or to a plasma with an ultra-relativistic temperature where $\Gamma=4/3$. However, in the high-temperature limit, i.e., $\Theta \to \infty$ with $\Gamma > 4/3$, the sound speed exceeds relativistic limit ($c_{s} >1/\sqrt{3}$). 
Moreover a constant $\Gamma$-law EoS is not consistent with relativistic kinetic theory, the so-called Taub's fundamental inequality, which requires the specific enthalpy to satisfy
\begin{equation} 
(h-\Theta)(h-4\Theta) \ge 1.
\end{equation} 
This rules out a constant $\Gamma$-law EoS with $\Gamma > 4/3$, if applied to $0 < \Theta < \infty$.

The theory of single-component perfect gases in the relativistic regime shows that the specific enthalpy is a function of the temperature $\Theta =p/\rho$ only, and has the form (Synge 1957)
\begin{equation}  
h = {K_{3} (1/\Theta) \over K_{2} (1/\Theta)},
\end{equation}
where $K_{2}$ and $K_{3}$ are the $2^{nd}$ and $3^{rd}$ order modified Bessel functions of the second kind respectively.
Using an equivalent $\Gamma_{eq} =  (h-1)/( h-1-\Theta)$ in the non-relativistic temperature limit ($\Theta \to 0$) yields $\Gamma_{eq} \to 5/3$, and in the ultra-relativistic temperature limit ($\Theta \to \infty$) yields $\Gamma_{eq} \to 4/3 $ (see Fig. 1a). However, this EoS requires extra computational costs because the thermodynamics of the fluid is expressed in terms of the modified Bessel functions (Falle \& Komissarov 1996).

Recently, Mignone et al. (2005) proposed an EoS, the so-called TM EoS, that follows eq. (31) well. The TM EoS, which was first introduced by Mathews (1971), is given by
\begin{equation} 
p={\rho \epsilon (\rho \epsilon + 2 \rho) \over 3 (\rho \epsilon + \rho)} \mbox{  or  }  
h={5 \over 2} \Theta + \sqrt{{9 \over 4} \Theta^{2} +1},
\end{equation}
and the sound speed is calculated from
\begin{equation}
c_{s}^{2} = {\Theta \over 3h} {5h - 8 \Theta \over h-\Theta}.
\end{equation}
The TM EoS corresponds to the lower bound of Taub's fundamental inequality, i.e., $(h-\Theta)(h-4\Theta)=1$, and produces the correct asymptotic values for $\Gamma_{eq}$.

Ryu et al. (2006) proposed an EoS which is a simpler algebraic function of $\Theta$,  hereafter referred to as the RC EoS, that satisfies Taub's inequality for all $\Theta$. The RC EoS is given by
\begin{equation}
{p \over \epsilon - p} = {3p + 2 \rho \over 9 p + 3 \rho} \mbox{ or } 
h=2{6 \Theta^{2} + 4 \Theta + 1 \over 3 \Theta +2},
\end{equation}
and the sound speed is calculated from
\begin{equation}
c_{s}^{2} = {\Theta (3\Theta +2)(18 \Theta^{2} + 24 \Theta + 5) \over 3 (6 \Theta^{2} + 4 \Theta +1)(9 \Theta^{2}+ 12 \Theta +2)}.
\end{equation}

\begin{figure}[h!]
\epsscale{1.1}
\plotone{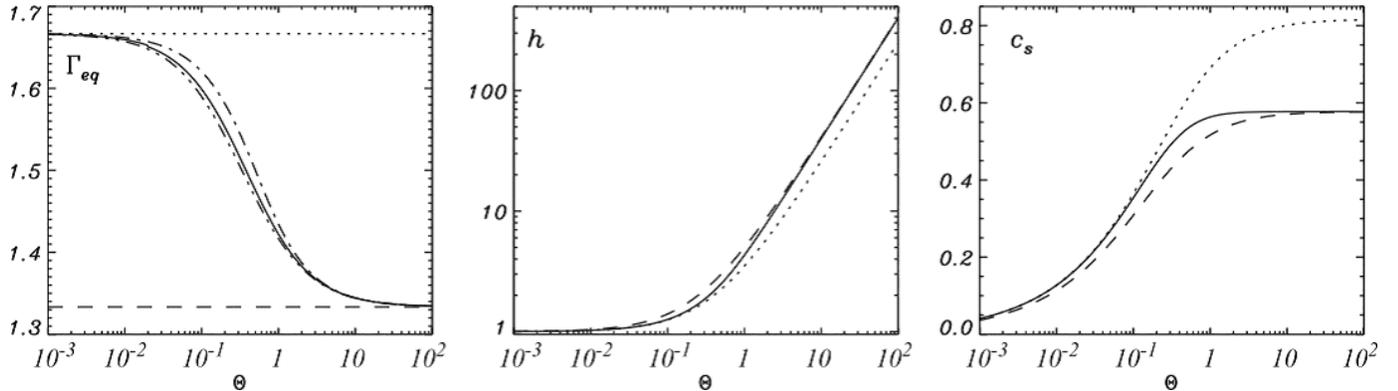}
\caption{Equivalent $\Gamma$ (left), specific enthalpy (middle), and sound speed (right) as functions of the temperature $\Theta=p/\rho$. Different lines correspond to: constant $\Gamma$-law EoS with $\Gamma=5/3$ (dotted lines), constant $\Gamma$-law EoS with $\Gamma=4/3$ (dashed lines), TM EoS (dash-dotted lines) and RC EoS (dash-two dotted lines). For comparison Synge's EoS has been plotted as the solid lines. 
\label{f1}}
\end{figure}
Figure 1 shows the equivalent $\Gamma$, the specific enthalpy and the sound speed as a function of $\Theta$ for TM EoS, RC EoS, Synge's EoS as well as a constant $\Gamma$-law EoS with $\Gamma=5/3$ and $4/3$. 
The specific enthalpy and the sound speed computed using the TM EoS and the RC EoS are well matched to Synge's EoS and cannot be distinguished on the plots. The approximations to Synge's EoS such as the TM EoS and RC EoS are hereafter referred to as approximate EoSs.  

\section{Numerical Method}

A well-known and challenging feature of the system of equations (6)-(13) is that they have source terms for the evolution of the electric field that become stiff in the high conductivity (low resistivity) limit. Following Komissarov (2007), we will use the Strang-splitting technique (Strang 1968). 

The system of equations (6)-(13) can be written as a single phase vector equation
\begin{equation} 
{\partial U(P) \over \partial t } + {\partial F^{m} (P) \over \partial x^{m}} = S(P),
\end{equation}
where
\vspace{0.1cm}
\[ 
U = \left( \begin{array}{c}
D \\
m^{i} \\
\tau \\
E^{i} \\
B^{i} \\
\Psi \\
\Phi \\
q
\end{array} \right), \ 
P=\left( \begin{array}{c}
\rho \\
v^{i} \\
p \\
E^{i} \\
B^{i} \\
\Psi \\
\Phi \\
q
\end{array} \right), \ 
S=\left( \begin{array}{c}
0 \\
0^{i} \\
0 \\
-j^{i} \\
0^{i} \\
q-\kappa \Psi \\
-\kappa \Phi \\
0 
\end{array} \right)
\]
are the vectors of conserved variables, primitive variables and sources, respectively, and
\vspace{0.1cm}
\[ 
F^{m} = \left( \begin{array}{c}
\rho \gamma v^{m} \\
\Pi^{im} \\
Y^{m} \\
-e^{imk} B_{k} + \Psi g^{im} \\
e^{imk} E_{k} + \Phi g^{im} \\
E^{m} \\
B^{m} \\
j^{m}
\end{array} \right)
\]
\vspace{0.1cm}
is the vector of numerical fluxes, where $e^{ijk}$ is the Levi-Civita alternating tensor of space. 
 
 The source term can be split into two parts
\[
 S_{a}(P)=\left( \begin{array}{c}
0 \\
0^{i} \\
0 \\
-qv^{i} \\
0^{i} \\
q \\
0 \\
0
\end{array} \right) \mbox{and} \ S_{b}(P)=\left( \begin{array}{c}
0 \\
0^{i} \\
0 \\
-j_{c}^{i} \\
0^{i} \\
-\kappa \Psi \\
-\kappa \Phi \\
0
\end{array} \right),
\]
where
\begin{equation}
\vec{j}_{c} = \sigma \gamma [\vec{E} + \vec{v} \times \vec{B} - (\vec{E} \cdot \vec{v}) \vec{v}]
\end{equation}
is the conductivity current. The source term $S_{b}$ is a stiff relaxation term that requires special care to capture the dynamics in a stable and accurate manner. In the Strang time-step splitting technique, firstly the solution is advanced using the stiff-part equations
\begin{equation}
{\partial U(P) \over \partial t} = S_{b}(P)
\end{equation}
over the half time-step, $\Delta t/2$. Secondly, advance of the non-stiff part of the equations is made via second-order accurate numerical integration of
\begin{equation}
{\partial U(P) \over \partial t} + {\partial F^{m} (P) \over \partial x^{m}}= S_{a}(P)
\end{equation} 
over the full time step. Thirdly, again the solution is advanced by the stiff-part equations over the half-time step.

Time advance of the non-stiff part of the equations is given by
\begin{equation}
U_{n+1} = U_{n} + \Delta t \sum^{n_d}_{m=1} {F_{m-1/2, n+1/2} - F_{m+1/2, n+1/2} \over \Delta x^{m}} + \Delta t S_{a, n+1/2},
\end{equation} 
where $U_{n}$ represents the cell-centered conserved variables at $t=t_{n}$, $U_{n+1}$ represents the cell-centered conserved variables at $t=t_{n}+\Delta t$, $S_{a, n+1/2}$ represents the cell-centered non-stiff source term at $t=t_{n}+\Delta t/2$, $F_{m+1/2, n+1/2}$ is the numerical flux though the right-hand side cell interface and $F_{m-1/2, n+1/2}$ is the numerical flux though the left-hand side cell interface in the direction of $x^{m}$ at time $t=t_{n}+\Delta t/2$, $\Delta x^{m}$ is the cell size in this direction, and $n_{d}$ is the number of spatial dimensions. The non-stiff sources and numerical fluxes at the half-time step are calculated from
\begin{equation}
U_{n+1/2} = U_{n} + {\Delta t \over 2} \sum^{n_d}_{m=1} {F_{m-1/2, n} - F_{m+1/2, n} \over \Delta x^{m}} + {\Delta t \over 2} S_{a, n}.
\end{equation} 
The numerical fluxes at the cell-interface $F_{m+1/2, n}$ are calculated using the simplified Harten-Lax-van Leer (HLL) approximate Riemann solver (Harten et al. 1983, Komissarov 2007) where the maximum characteristic speed of the system in each direction equals the speed of light. The left-hand and right-hand states of each cell interface using the HLL approximate Riemann solver are computed from various reconstruction schemes as in our ideal RMHD code (Mizuno et al. 2006; 2011). In this paper, we use a piecewise linear method (PLM) reconstruction such as the minmod slope-limited linear interpolation scheme or the Monotonized Central (MC) slope-limited linear interpolation scheme as these are the simplest reconstruction schemes that capture a shock sharply. The resulting scheme is second-order accurate in time and space.

Following the work of Komissarov (2007), the split evolution equations of the electric field, eqs. (23) \& (24), can be solved analytically
\begin{eqnarray}
\vec{E}_{\parallel} &=& \vec{E}_{\parallel}^{0} \exp \left[ - {\sigma \over \gamma} t \right], \\
\vec{E}_{\perp} &=& \vec{E}_{\perp}^{*} + (\vec{E}_{\perp}^{0} - \vec{E}_{\perp}^{*}) \exp [-\sigma \gamma t],
\end{eqnarray}
where $E_{\perp}^{*} = - \vec{v} \times \vec{B}$ and suffix 0 indicates the initial component. The stiff-part equations related to the two scalar fields $\Psi$ and $\Phi$ are also solved analytically with solution
\begin{eqnarray} 
\Psi &=& \Psi_{0} \exp [-\kappa t], \\
\Phi &=& \Phi_{0} \exp [-\kappa t],
\end{eqnarray}
where $\Psi_{0}$ and $\Phi_{0}$ are the initial values of $\Psi$ and $\Phi$.

In order to evolve this system of equations, the numerical fluxes $F^{m}$ must be computed at each time-step. These fluxes depend on the primitive variables $P$, which must be recovered from the evolved conserved variables $U$.
In conserved variables, $E$ and $B$ can be calculated by evolving Maxwell's equations. However, it is more stable to evolve the stiff part equations (42) - (43) during the primitive recovery process when $\sigma$ becomes large (Palenzuela et al. 2009). The primitive recovery procedure adapted to an approximate EoS such as the TM EoS and the RC EoS follows that used by Palenzuela et al. (2009).   

\section{Numerical Tests}

In this section, one-dimensional and two-dimensional tests are presented. Three one-dimensional tests have been used to validate our new resistive relativistic MHD code in different regimes. A one-dimensional shock-tube test and three two-dimensional tests have been used to investigate the effect of different EoSs. The damping coefficient of the hyperbolic divergence cleaning is set to $\kappa=1$.  The magnetic field is divergence-free and charge is preserved at the truncation error level.

\subsection{One-dimensional tests}

\subsubsection{Large amplitude CP Alfv\'{e}n wave test}

This test consists of the propagation of a large amplitude circularly-polarized Alfv\'{e}n wave along a uniform back-ground magnetic field $B_{0}$ in a domain with periodic boundary conditions. The exact solution is given by Del Zanna et al. (2007) in the ideal MHD limit, and was used as an ideal-MHD limit test problem by Palenzuela et al. (2009) and Takamoto \& Inoue (2011). Here we use conditions  similar to previous studies with
\begin{eqnarray}
(B_{y}, B_{z}) &=& \zeta_{A} B_{0} (\cos [k(x-v_{A}t], \sin [k(x-v_{A})t]), \\
(v_{y}, v_{z}) &=& - {v_{A} \over B_{0}} (B_{y}, B_{z}),
\end{eqnarray}
where $B_{x}=B_{0}$, $v_{x}=0 $, $k$ is the wave number and $\zeta_{A}$ is the amplitude of the wave.
The special relativistic Alfv\'{e}n speed $v_{A}$ is given by 
\begin{equation}
v_{A}^{2} = {2B_{0}^{2} \over h+ B_{0}^{2}(1+ \zeta_{A}^{2})} \left(   1+ \sqrt{ 1- \left( {2 \zeta_{A} B_{0}^{2} \over h + B_{0}^{2} (1 + \zeta_{A}^{2})}\right) } \right)^{-1}.
\end{equation}

For this test we use initial parameters  $\rho=p=1$ and $B_{0}=0.46188$, the Alfv\'{e}n velocity $v_{A}=0.25c$, and we adopt a constant gamma-law EoS with $\Gamma=2$. Following Palenzuela et al. (2009), we use a high uniform conductivity $\sigma=10^{5}$ with three different resolutions of 50, 100 and 200 cells covering the computational domain $ x \in [-0.5, 0.5]$. Figure 2 shows the numerical results at $t=4$ (one Alfv\'{e}n crossing time) for the three different resolutions.  This result shows that the new resistive RMHD code reproduces ideal relativistic MHD solutions when the conductivity $\sigma$ is high.
\begin{figure}[h!]
\epsscale{0.7}
\plotone{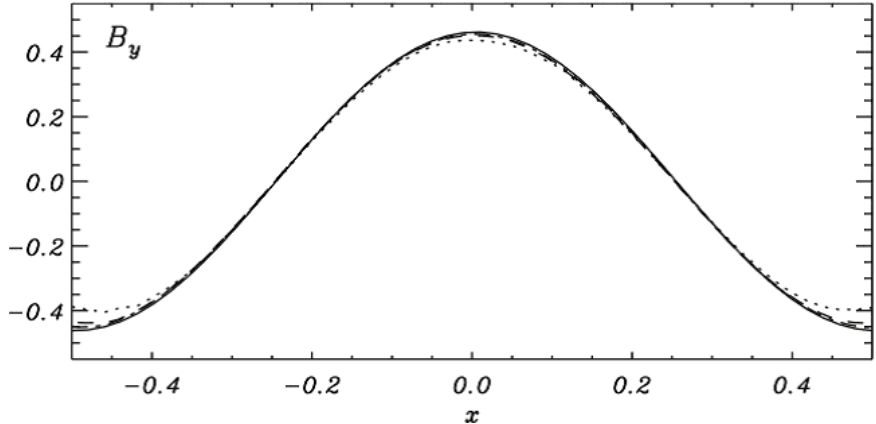}
\caption{Magnetic field component $B_{y}$ in a large-amplitude CP Alfv\'{e}n wave test using three different resolutions $N=50$ (dotted), $100$ (dashed) and $200$ (dash-dotted) at $t=4$. The solid line shows the exact solution. The numerical results are in good agreement with the analytical one (the highest resolution is excellent). \label{f2}}
\end{figure}
The $L_{1}$ norm errors of the magnetic field component $B_{y}$ in this test are shown in Figure 3. The numerical result is slightly shallower than $2^{nd}$ order convergence. This is likely caused by the periodic boundary. 
\begin{figure}[h!]
\epsscale{0.7}
\plotone{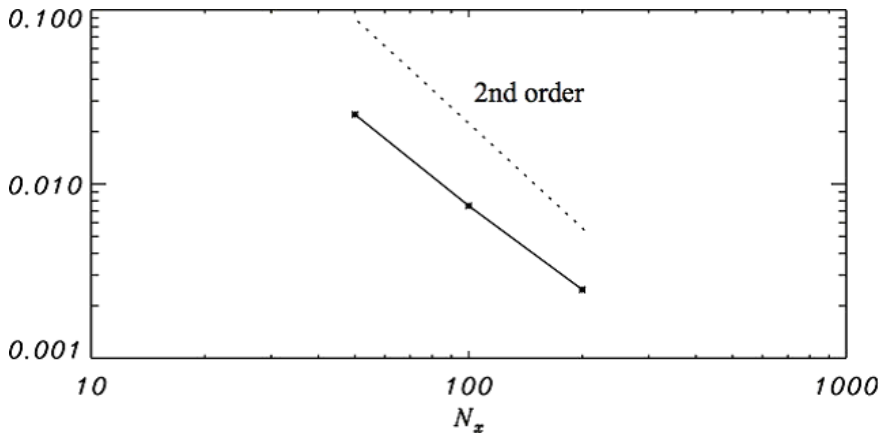}
\caption{$L_{1}$ norm errors of the magnetic field component $B_{y}$ in a large-amplitude CP Alfv\'{e}n wave test using the three different resolutions $N=50$, $100$ and $200$. \label{f3}}
\end{figure}

\subsubsection{1D self-similar current sheet test}

This test problem has been used for moderate resistivity cases (e.g., Komissarov 2007; Palenzuela et al. 2009; Takamoto \& Inoue 2011). In this test problem the magnetic pressure is much smaller than the gas pressure everywhere. The magnetic field configuration is given by $B=[0, B_{y}(x,t), 0]$, where $B_{y}(x,t)$ changes sign within a thin current layer of thickness $\Delta l$. An initial solution is provided in equilibrium with $p= $ constant. The evolution of this thin current layer is a slow diffusive expansion due to the resistivity and described by the diffusion equation
\begin{equation}
\partial_{t} B_{y} - {1 \over \sigma } \partial^{2}_{x} B_{y} =0.
\end{equation}
As the thickness of the layer becomes much larger than $\Delta l$ the expansion becomes self-similar with
\begin{equation}
B_{y}(x,t) = B_{0}\  \mbox{erf} \left( {1 \over 2} \sqrt{\sigma \over \chi} \right),
\end{equation}
where $\chi = t/x^{2}$ and erf is the error function. This analytic result can be used for testing the moderate resistivity regime. 

In the test problem, we have chosen an initial solution at $t=1$ with $p=50$, $\rho=1$, $\vec{E} = \vec{v}=0$ and $\sigma=100$ ($\eta=1/\sigma=0.01$). A constant gamma-law EoS with $\Gamma=2$ is used. The computational domain is uniform with 200 cells in $[-1.5, 1.5]$. 
\begin{figure}[h!]
\epsscale{0.7}
\plotone{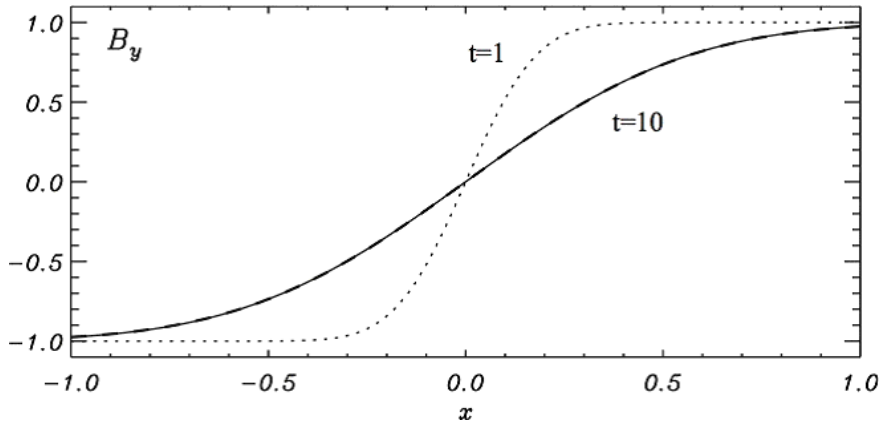}
\caption{Magnetic field component $B_{y}$ in a Self-similar current sheet test. The dotted and dashed lines are indicated the analytical solution at t=1 and t=10. The solid line shows the numerical solution at $t=10$. The numerical solution is in excellent agreement with the analytical one. \label{f4}}
\end{figure}
The numerical simulation is evolved up to $t=10$ and then the numerical solution is compared in Figure 4 with the analytical solution. The numerical and analytical solutions cannot be distinguished on the plot. This indicates that the moderate resistivity regime is well described by the code. 

\subsubsection{1D shock-tube tests}

As a first shock-tube test in the restive relativistic MHD regime, we consider a simple MHD version of the Brio and Wu test as in Palenzuela et al. (2009) and Takamoto \& Inoue (2011). The initial left and right states are separated at $x=0.5$ and are given by
\begin{eqnarray}
(\rho^{L}, p^{L}, B^{L}_{y}) &=& (1.0, 1.0, 0.5) \\
(\rho^{R}, p^{R}, B^{R}_{y}) &=& (0.125, 0.1, -0.5).
\end{eqnarray}
All other fields are set to $0$. We use a constant $\Gamma$-law EoS with $\Gamma=2$. The computational domain covers the region $x \in [-0.5,0.5]$ with $200$ cells.

Figure 5 shows the numerical results at $t=0.4$ for conductivities $\sigma=0$, $10$, $10^2$, $10^3$, $10^5$. 
The exact solution to the ideal RMHD Riemann problem was found by Giacomazzo \& Rezzolla (2006). When $B_{x}=0$, the solution contains only two fast waves, a left-moving rarefaction wave and a right-moving shock with a tangential discontinuity between them.
\begin{figure}[h!]
\epsscale{0.6}
\plotone{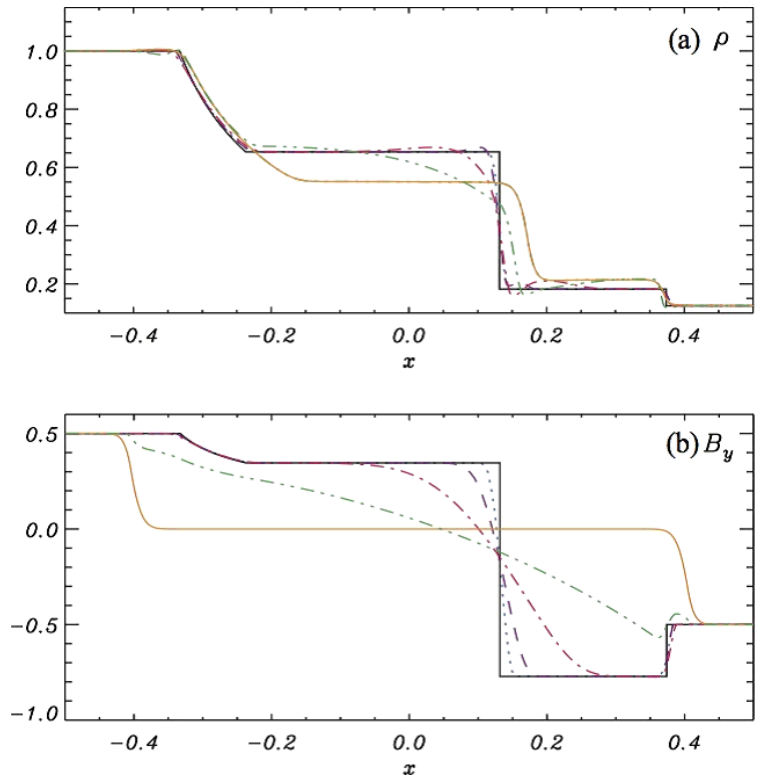}
\caption{(a) Density and (b) magnetic field component $B_{y}$ in the simplified Brio \& Wu shock-tube test. Different lines indicate different conductivities: $\sigma=0$ (orange solid), $10$ (green dash-two-dotted), $10^2$ (red dash-dotted), $10^3$ (purple dashed), $10^5$ (blue dotted). The black solid line shows the exact solution in the ideal RMHD case. \label{f5}}
\end{figure}
The results show that the solution smoothly changes from a wave-like solution for $\sigma=0$ to the ideal-MHD solution for high conductivity $\sigma=10^5$. Note that for $\sigma=0$ the solution describes a discontinuity propagating at the speed of light corresponding to Maxwell's equations in vacuum. This result is nearly the same as test results for other codes (Palenzuela et al. 2009; Takamoto \& Inoue 2011).

Palenzuela et al. (2009) reported that Strang's splitting technique became unstable for moderately high conductivity in this shock tube test, and they suggested using an implicit method. However, Takamoto \& Inoue (2011) found that this instability is not related to Strang's splitting technique but instead to the calculation of the electric field during the primitive recovery procedure. In this new resistive RMHD code, the shock tube test is solved stably using Strang's splitting technique even when $\sigma \simeq 10^{6}$.    

Balsara Test 2 (Balsara 2001) is used as a second shock tube test to investigate the effect of different EoSs in the resistive relativistic MHD regime. In ideal RMHD, Mignone \& McKinney (2007) have already performed this test to check the effect of different EoSs. In this test the initial left and right states are separated at $x=0.5$ and are given by
\begin{eqnarray}
(\rho^{L}, p^{L}, B^{L}_{x}, B^{L}_{y}, B^{L}_{z}) &=& (1.0, 30.0, 5.0, 6.0, 6.0) \\
(\rho^{R}, p^{R}, B^{R}_{x}, B^{R}_{y}, B^{R}_{z}) &=& (1.0, 1.0, 5.0, 0.7, 0.7).
\end{eqnarray}
The computational domain covers the region $x \in [-0.5,0.5]$ with $800$ cells. This test shows that a mildly relativistic blast wave propagates to the right with maximum Lorentz factor of $1.3 \le \gamma \le 1.4$. 

The numerical results at $t=0.4$ for the constant $\Gamma$-law EoS with $\Gamma=5/3$, the TM EoS and the RC EoS using conductivities $\sigma=0$, $10$, $10^2$, $10^3$ are shown in Figures 6, 7 and 8, respectively. 
\begin{figure}[h!]
\epsscale{1.0}
\plotone{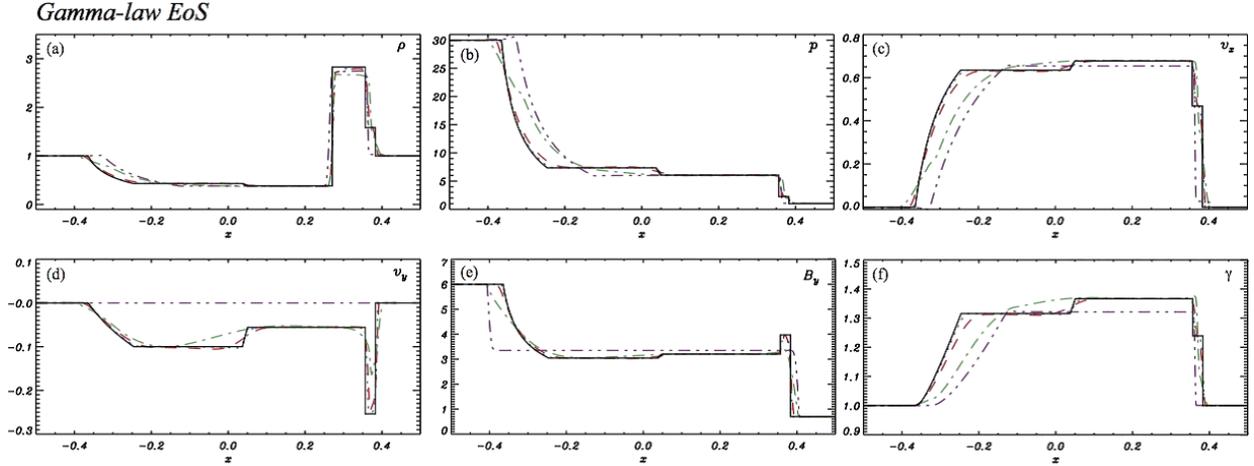}
\caption{(a) Density, (b) gas pressure, (c) velocity component $v_{x}$, (d) velocity component $v_{y}$, (e) magnetic field component $B_{y}$ and (f) Lorentz factor in the Blasara Test 2 (mildly relativistic blast wave) at $t=0.4$ using an ideal EoS with $\Gamma=5/3$. Different lines indicate different conductivity: $\sigma=0$ (purple dash-two-dotted), $10$ (green dash-dotted), $10^2$ (red dashed), $10^3$ (blue dotted). The black solid line shows the exact solution in the ideal RMHD case. \label{f6}}
\end{figure}
\begin{figure}[h!]
\epsscale{1.0}
\plotone{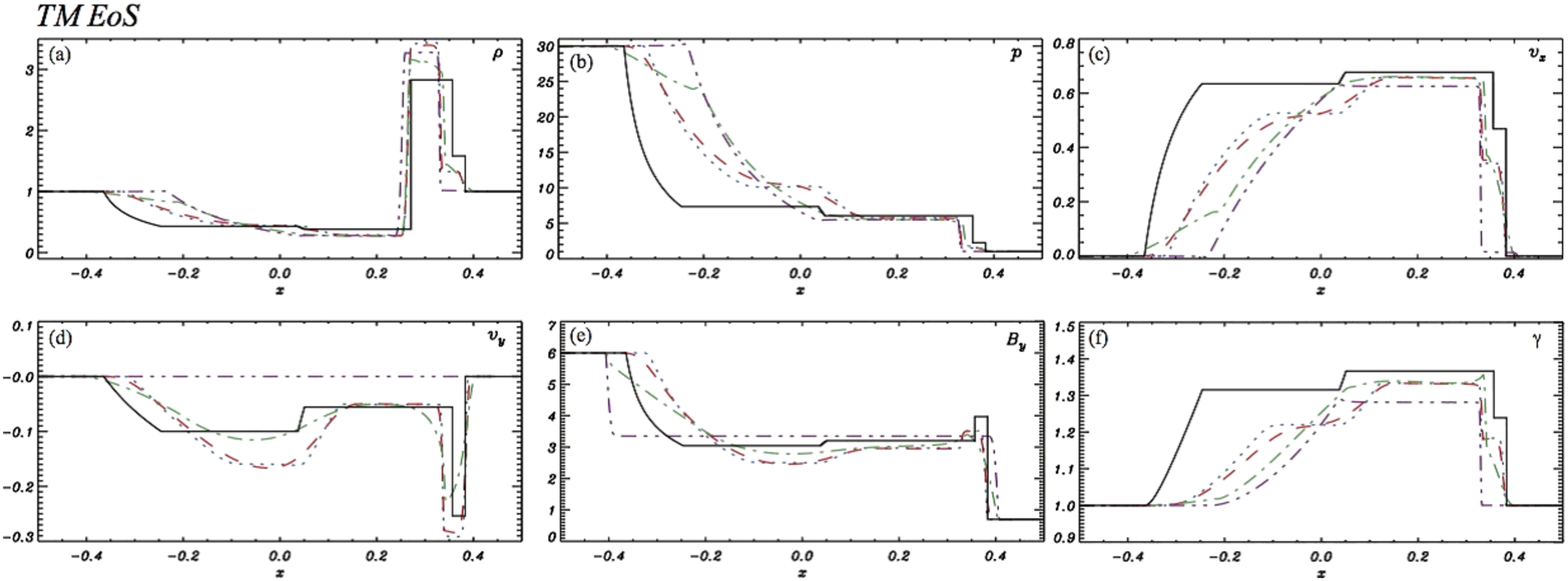}
\caption{ The same as in Fig. 6 but using the TM EoS. \label{f7}}
\end{figure}
\begin{figure}[h!]
\epsscale{1.0}
\plotone{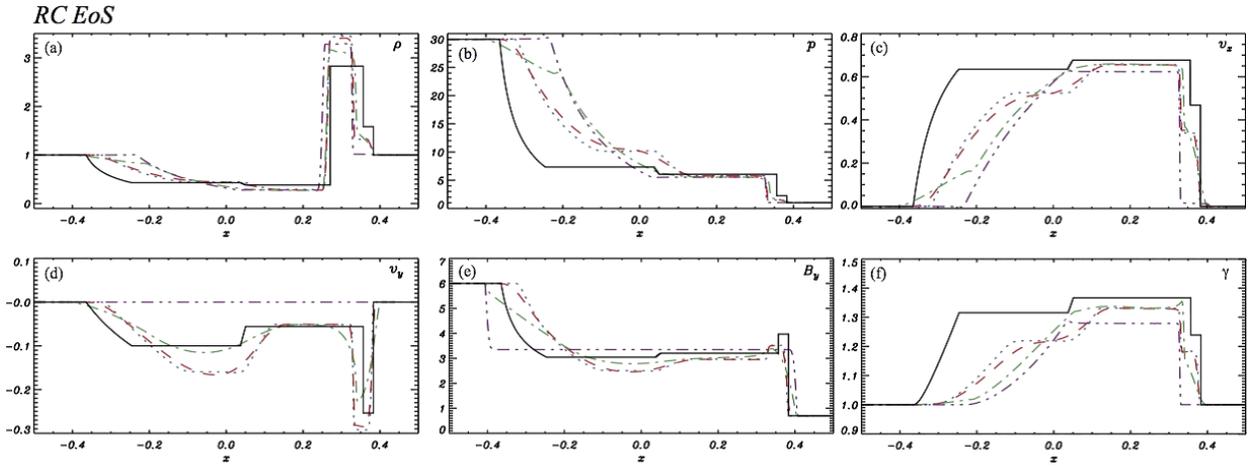}
\caption{ The same as in Fig. 6 but using the RC EoS.  \label{f8}}
\end{figure}
The solutions show fast and slow rarefaction waves, a contact discontinuity, and slow and fast shocks from left to right. In these cases, no rotational discontinuity is seen. The results obtained from the TM EoS and RC EoS cases are considerably different from the constant $\Gamma$-law EoS with $\Gamma=5/3$. In the approximate EoS cases, the rarefaction waves and shocks propagate with smaller velocities. This is predicted from the lower sound speed in the approximate EoS cases relative to overestimated sound speed in the ideal EoS case with $\Gamma=5/3$. Behind the slow shock, the approximate EoS cases have a higher peak density, which follows from the previous considerations. These properties are consistent with those in the ideal RMHD case (Mignone \& McKinney 2007).  On the other hand, the results obtained from TM EoS and RC EoS cases are very similar at our numerical resolution. This similarity reflects the similarity in the distributions of specific enthalpy (see Fig. 1).     
Again the results show a smooth change from a wave-like solution for $\sigma=0$ towards  an ideal-MHD solution for the highest conductivity, $\sigma=10^3$, for all the different EoS cases.
The differences between the approximate EoS cases and the constant $\Gamma$-law case are larger at higher $\sigma$ where the approximate EoS cases approach the ideal MHD case more slowly. Even for low conductivity, i.e, $\sigma=10$, we clearly see a difference between the ideal EoS case with $\Gamma=5/3$ and the approximate EoS cases.

\subsection{Two-dimensional tests}

\subsubsection{The Cylindrical Explosion}
  
 We now consider tests involving shocks in multi-dimensions. Firstly we choose a test involving a cylindrical blast wave expanding into an initially uniform magnetic field. This is a standard test for ideal relativistic MHD codes even though there is no exact solution because this test will reveal subtle bugs and potential weaknesses in the numerical implementation. For this test, we use a Cartesian computational domain $(x,y) \in [-6,6]$ with 200 uniform cells in each direction. The initial explosion is initialized  by setting the gas pressure and density to $p=1$ and $\rho=0.01$ within a cylinder of radius $r < 0.8$ centered on the origin. In an intermediate region $0.8 < r < 1.0$, the pressure and density decrease exponentially to that of the ambient gas which has $p=\rho=0.001$.  The initial magnetic field is uniform in the $x$-direction with $\vec{B}=(0.05, 0, 0)$. The other quantities are set to zero (i.e., $\vec{v}=\vec{E}=q=0$). 
   
Figure 9 shows the magnetic field components $B_{x}$ and $B_{y}$ at $t=4$ using the ideal EoS with $\Gamma=4/3$ and the TM EoS. 
\begin{figure}[h!]
\epsscale{1.0}
\plotone{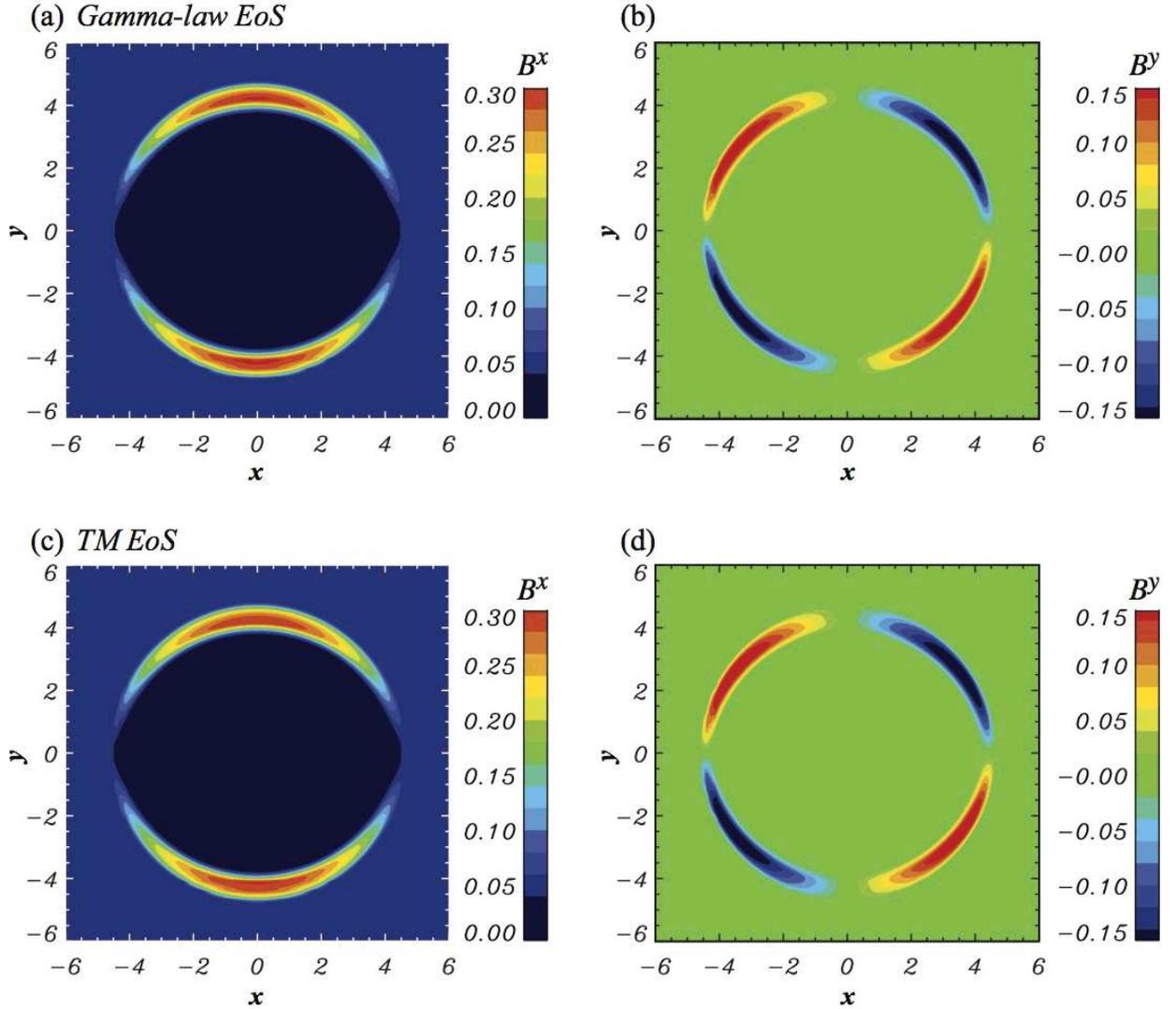}
\caption{ Magnetic field components $B_{x}$ (left panels) and $B_{y}$ (right panels) for the cylindrical explosion test at $t=4$ using a ideal EoS with $\Gamma=4/3$ (upper panels) and TM EoS (lower panels).  \label{f9}}
\end{figure}
The ideal-MHD simulation is performed using a high conductivity of $\sigma=10^5$. 
The results are qualitatively similar to those reported in previous studies in ideal RMHD (Komissarov 1999; Leismann et al. 2005; Neilsen et al. 2006; Del Zanna et al. 2007; Mizuno et al. 2011) and RRMHD (Komissarov 2007; Palenzuela et al. 2009). The results obtained from the ideal EoS with $\Gamma=4/3$ and the TM EoS are qualitatively very similar. This means that an ideal EoS with $\Gamma=4/3$ satisfactorily captures all the shock properties.

Figure 10 shows one-dimensional profiles of the gas and magnetic pressure along the $y$-axis for
\begin{figure}[h!]
\epsscale{0.9}
\plotone{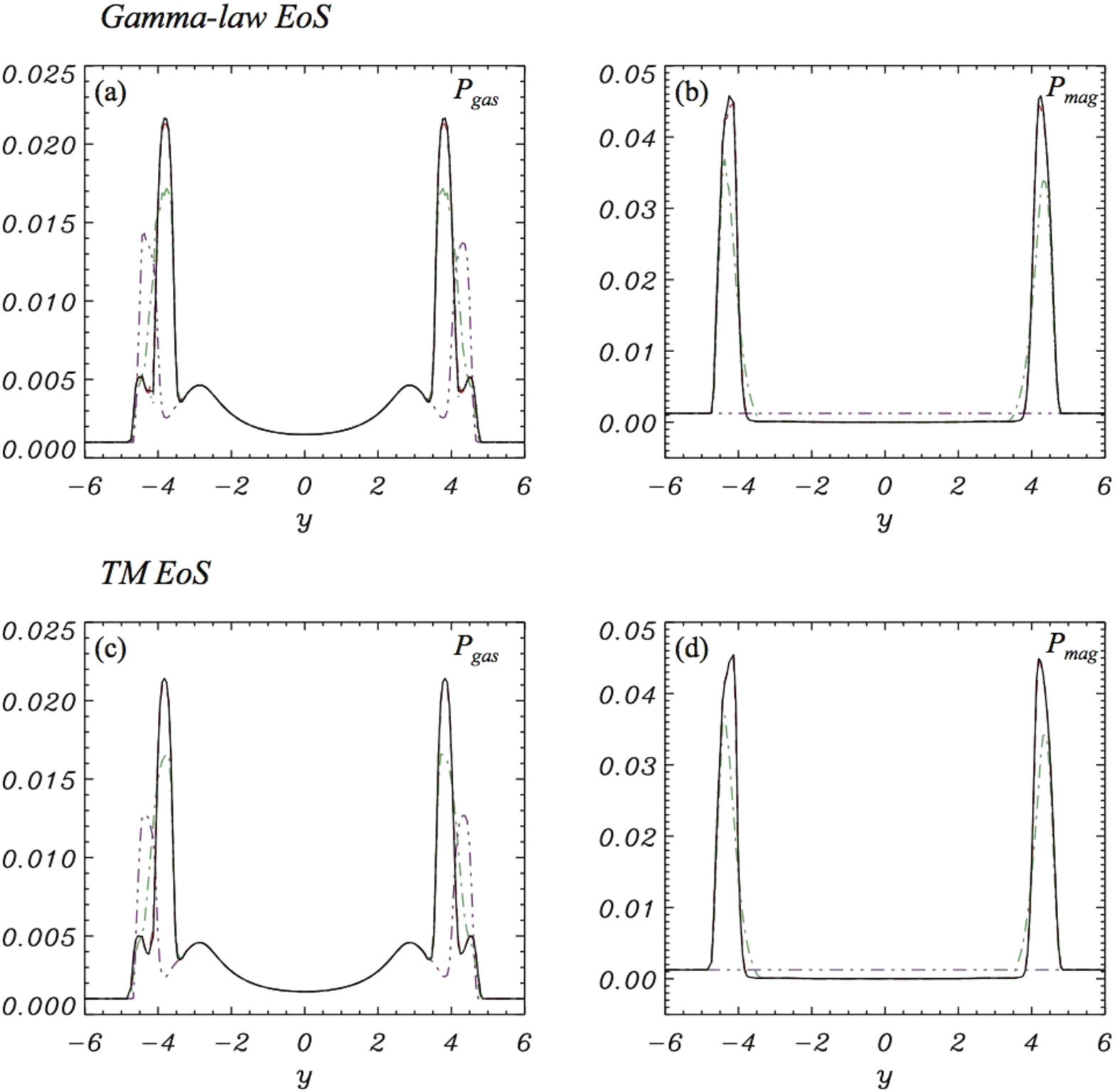}
\caption{Gas pressure $P_{gas}$ (left panels) and magnetic pressure $P_{mag}$ (right panels) for the cylindrical explosion test at $t=4$ using an ideal EoS with $\Gamma=4/3$ (upper panels) and a TM EoS (lower panels). The different lines show conductivity cases: $\sigma=0$ (purple dash-two-dotted), $10$ (green dash-dotted), $10^2$ (red dashed), $10^3$ (blue dotted), and $10^5$ (black solid). \label{f10}}
\end{figure}
conductivities of $\sigma=0$, $10$, $10^2$, $10^3$, $10^5$ using the ideal EoS with $\Gamma=4/3$ and the TM EoS.  At high conductivity $\sigma \ge 10^3$, we do not see any significant difference. This means that high conductivity recovers the ideal-MHD solution. As the conductivity decreases, the maximum gas and magnetic pressure decrease. Of course there is no magnetic pressure increase for $\sigma=0$. 

\subsubsection{Kelvin-Helmholtz instability test}

We present calculations of the linear and nonlinear growth of the two-dimensional Kelvin-Helmhotz instability (KHI) to investigate the effect of conductivity and the EoS on the development of turbulence in the resistive relativistic MHD regime.

Initial conditions for this test are taken from a combination of previous studies in ideal RMHD (Bucciantini \& Del Zanna 2006; Mignone et al. 2009; Beckwith \& Stone 2011). The shear velocity profile is given by
\begin{equation}
 v_{x}=\left\{ \begin{array}{cl}
v_{sh} \tanh \left( {y-0.5 \over a} \right)  & \mbox{if $ y > 0.0$} \\
-v_{sh} \tanh \left( {y+0.5 \over a} \right) & \mbox{if $ y \le 0.0$}
\end{array} \right.
\end{equation}  
Here, $a=0.01$ is the characteristic thickness of the shear layer, and $v_{sh}=0.5$ corresponds to a relative Lorentz factor of $2.29$. The initial uniform pressure is $p=1.0$. The density is initialized using the shear velocity profile, with $\rho=1.0$ in regions with $v_{sh}=0.5$ and $\rho=10^{-2}$ in regions with $v_{sh}=-0.5$. The magnetic field components are given in terms of the poloidal and toroidal magnetization parameters $\mu_{p}$ and $\mu_{t}$ as
\begin{equation}
(B_{x}, B_{y}, B_{z})=(\sqrt{2 \mu_{p} p}, 0, \sqrt{2 \mu_{t} p}),
\end{equation}
with $\mu_{p}=0.01$ and $\mu_{t}=1.0$. The instability is seeded by a single mode perturbation of the form
\begin{equation}
 v_{y}=\left\{ \begin{array}{cl}
A_{0} v_{sh} \sin(2 \pi x) \exp \left[ -\left( {y-0.5 \over \alpha} \right)^{2} \right]  & \mbox{if $ y > 0.0$} \\
- A_{0} v_{sh} \sin(2 \pi x) \exp \left[ -\left( {y+0.5 \over \alpha} \right)^{2} \right] & \mbox{if $ y \le 0.0$}
\end{array} \right.
\end{equation}  
Here, $A_{0}=0.1$ is the perturbation amplitude and $\alpha=0.1$ is the characteristic length scale over which the perturbation amplitude decreases exponentially. The computational domain covers $x \in [-0.5, 0.5]$, $y \in 
[-1,1]$ with $256 \times 512$ cells.

Figure 11 shows the perturbation amplitude  ($\Delta v_{y} \equiv (v_{y, max} - v_{y, min})/2$) and volume averaged poloidal magnetic field ($B_{pol}$ $=$ $\sqrt{B_{x}^{2}+B_{y}^{2}}$) as a function of time for conductivities of $\sigma=0$, $10$, $10^{2}$, $10^{3}$, $10^{5}$ using an ideal EoS with $\Gamma=4/3$ and using the TM EoS. All cases show an initial linear growth phase. Except for $\sigma=0$, the ideal EoS and the TM EoS have almost the same growth rate and the maximum amplitude is reached at $t\sim 2$. The maximum amplitude indicates the transition from the linear to the nonlinear phase. The $\sigma=0$ cases exhibit a lower growth rate and later transition to the nonlinear phase than the higher conductivity cases. Thus, differences in the conductivity and EoS do not affect the growth of KHI, except for $\sigma=0$. 
\begin{figure}[h!]
\epsscale{0.9}
\plotone{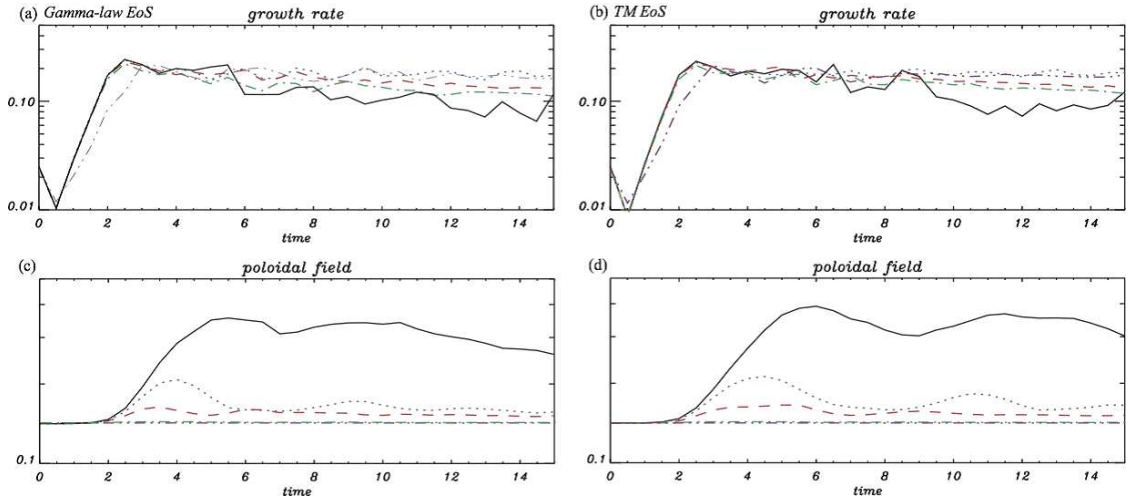}
\caption{ Evolution of the amplitude of the perturbation (upper panels) and volume-averaged poloidal field ($B_{pol}$) (lower panels) as a function of time for the Kelvin-Helmholtz instability test using a ideal EoS with $\Gamma=4/3$ (left panels) and TM EoS (right panels). The different lines indicate different conductivity cases: $\sigma=0$ (purple dash-two-dotted), $10$ (green dash-dotted), $10^2$ (red dashed), $10^3$ (blue dotted), $10^5$ (black solid). \label{f11}}
\end{figure}
  
Poloidal field amplification via stretching due to the main vortex developed by KHI follows the growth of KHI (see Fig. 12). In high conductivity cases, poloidal field amplification is very large, an increase by almost one-order of magnitude. Saturation in the poloidal field amplitude occurs latter than the transition from the linear to the non-linear KHI growth phase. This means that magnetic field amplification via stretching continues even after KHI is fully developed. Poloidal field amplification is weaker and saturation occurs earlier when the conductivity is low. Larger poloidal field amplification occurs for the TM EoS case than for the ideal EoS case. Therefore we find that magnetic field amplification via stretching due to the main vortex developed by KHI is strongly affected by the conductivity and the EoS.

Figures 12 and 13 show the time evolution of the density and the poloidal to toroidal field ratio ($B_{pol}/B_{tor}=\sqrt{B^{2}_{x} + B^{2}_{y}}/B_{z}$) for high conductivity, $\sigma=10^{5}$, using the ideal EoS with $\Gamma=4/3$ (Fig.\ 12) and using the TM EoS (Fig.\ 13). 
\begin{figure}[h!]
\epsscale{0.9}
\plotone{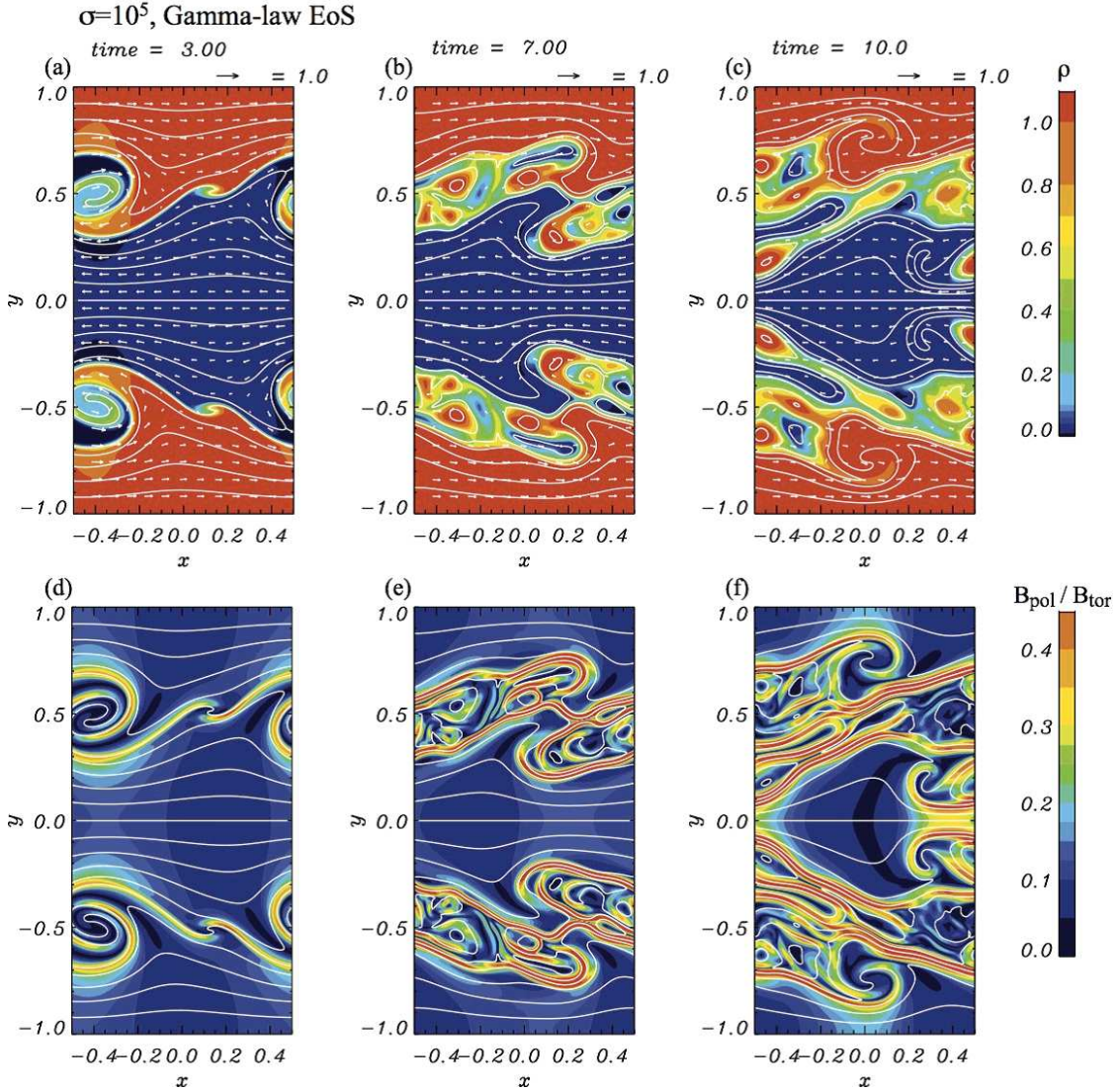}
\caption{Density (upper panels) and the poloidal to toroidal field ratio ($B_{pol}/B_{tor}$; lower panels) for the Kelvin-Helmholtz instability test at $t=3$, $7$, \& $10$ for $\sigma=10^{5}$ using the ideal EoS with $\Gamma=4/3$. White lines indicate magnetic field lines and the arrows show velocity vectors.\label{f12}}
\end{figure}
Both cases show formation of a main vortex by growth of KHI in the linear growth phase. In the ideal EoS case, a secondary vortex appears, although not fully developed. However, development of a secondary vortex is not found in the TM EoS case. Beckwith \& Stone (2011) found that a secondary vortex did not appear even in very high resolution simulations using the HLL approximate Reimann solver in ideal RMHD. In our simulations, we also used the HLL approximate Riemann solver to calculate the numerical flux, but do find a secondary vortex in the ideal EoS case. The difference is likely the result of the reconstruction scheme used here and the different reconstruction scheme used by Beckwith \& Stone. In the nonlinear phase the main vortex is distorted and stretched. The magnetic field is strongly amplified by shear motion in the vortex in the linear phase and by stretching in the nonlinear phase. As the mixing layer grows the field lines are bunched into a filamentary like stretched structure. In the TM EoS case, the vortex becomes strongly elongated along the flow direction. The structure created in the nonlinear phase is very different in the ideal EoS and the TM EoS cases.
\begin{figure}[h!]
\epsscale{0.9}
\plotone{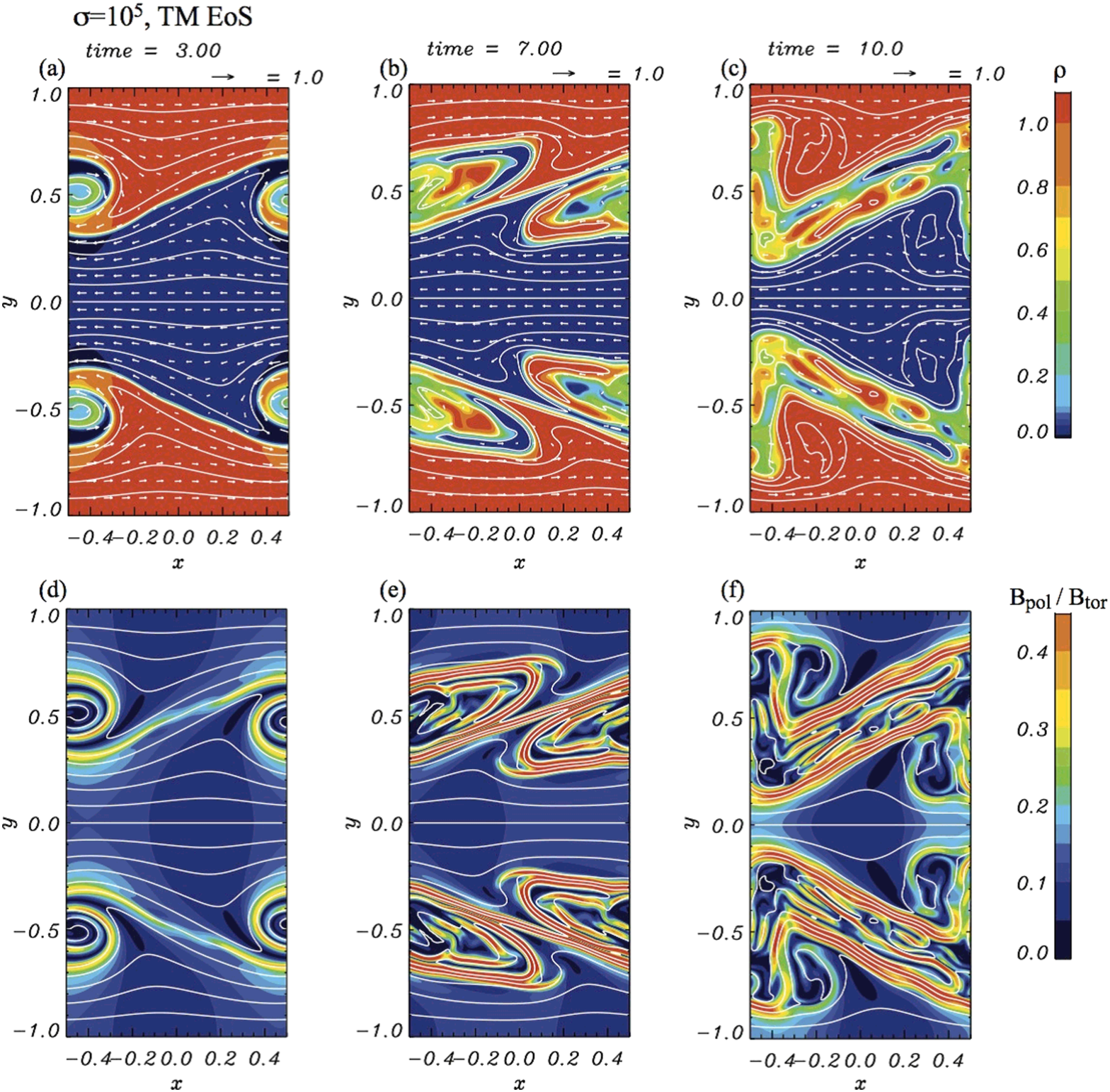}
\caption{ The same as Fig. 12 but using the TM EoS. \label{f13}}
\end{figure}

The field amplification structure for different conductivities from $\sigma=0$ to $10^{3}$ is shown in Figure 14. 
\begin{figure}[h!]
\epsscale{0.7}
\plotone{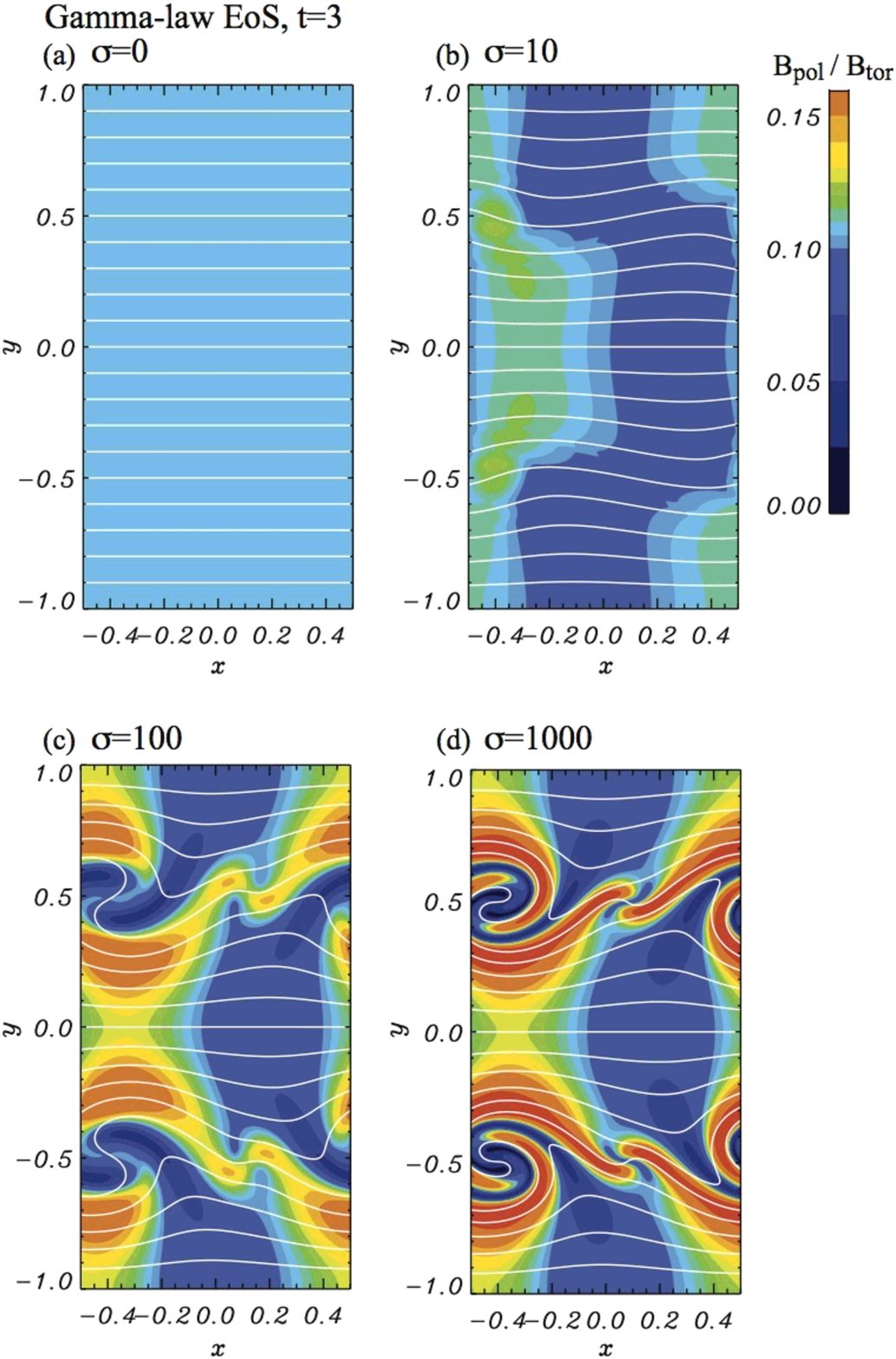}
\caption{ The poloidal to toroidal field ratio ($B_{pol}/B_{tor}$; lower panels) for the Kelvin-Helmholtz instability test at $t=3$ for (a) $\sigma=0$, (b) $\sigma=10$, (c) $\sigma=10^{2}$, and (d) $\sigma=10^3$ using the ideal EoS with $\Gamma=4/3$. \label{f14}}
\end{figure}
As seen in the time evolution of the averaged poloidal field shown in Fig.\ 11, the magnetic field amplification is weaker when the conductivity is low. Field amplification is a result of fluid motion in the vortex. In the high conductivity case, the magnetic field follows the fluid motion, like ideal MHD, and is strongly twisted. When the conductivity declines, the magnetic field is no longer strongly coupled to the fluid motion. Therefore the magnetic field is not strongly twisted. In the case using the TM EoS, we see the same trend for different conductivity and do not show the result here. 

\subsubsection{Relativistic Magnetic Reconnection test}

The final test involves relativistic magnetic reconnection. Pioneering work on relativistic magnetic reconnection using a resistive relativistic MHD code and 2D simulations was performed by Watanabe \& Yokoyama (2006) who considered Petschek-type reconnection. Zenitani et al. (2010) also have studied details of Petschek-type reconnection via 2D resistive relativistic MHD simulations. Zanotti \& Dumbser (2011) have investigated the dependence  of Petschek-type relativistic magnetic reconnection by performing 2D and 3D simulations over a broad range of conductivities and magnetizations. Takahashi et al. (2011) have studied Sweet-Parker type relativistic magnetic reconnection using 2D resistive relativistic MHD simulations. In this test, we present simulations of  Petschek-type reconnection and investigate the effect of the EoS.

We use initial conditions similar to that used in previous work (Watanabe \& Yokoyama 2006; Zenitani et al. 2010; Zanotti \& Dumbser 2011). The density and gas pressure are given by
\begin{eqnarray}
\rho &=& \rho_{b} + \mu_{m} \cosh^{-2} (y) \\
p &=& p_{b} + \mu_{m} \cosh^{-2} (y),
\end{eqnarray}
where $\rho_{b}=p_{b}=0.1$ are the uniform density and gas pressure outside the current sheet, and $\mu_{m}=B_{0}^{2}/(2 \gamma_{0}^{2})=1.0$ is the magnetization parameter. The velocity field is initially zero, hence $\gamma_{0}=1$. The magnetic field changes orientation across the current sheet according to
\begin{equation}
B_{x} = B_{0} \tanh (y),
\end{equation}   
where $B_{0}$ is calculated from the magnetization parameter. The current distribution is given by
\begin{equation}
j_{z} = B_{0} \cosh^{-2} (y).
\end{equation}
Over the whole computational domain there is a small background uniform resistivity $\eta_{b}=1/\sigma_{b}=10^{-3}$, except within a circle of radius $r_{\eta}=0.8$ which defines a region of anomalous resistivity with amplitude $\eta_{0}=1.0$, The resistivity can be written as 
\begin{equation}
\eta = \left\{ \begin{array}{cl}
\eta_{b} + \eta_{0}[2 (r/r_{\eta})^{3} - 3(r/r_{\eta})^{2} + 1] & \mbox{for $r \le r_{\eta}$,} \\
\eta_{b} & \mbox{for $r > r_{\eta}$,} 
\end{array} \right.
\end{equation}
where $r = \sqrt{x^{2} + y^{2}}$. The electric field is calculated from the resistivity distribution as 
\begin{equation}
E_{z} = \eta j_{z}.
\end{equation}
The computational domain is $x \in [-50, 50]$, $y \in [-20, 20]$ with $2000 \times 800$ cells, and outflow boundary conditions are used in both directions. 

Figure 15 shows the density, the $x$-component of the 4-velocity, $\gamma v_{x}$, and the out-of-plane current $j_{z}$ at $t=100$ using an ideal EoS with $\Gamma=4/3$ and the TM EoS. In this figure we confirm the essential features of  Petschek-type relativistic reconnection reported in previous work (Watanabe \& Yokoyama 2006; Zenitani et al. 2010; Zanotti \& Dumbser 2011). In both the ideal and TM EoS cases, we see similar time evolution and morphology. After an initial adjustment stage of $t \le 10$, the reconnection process starts around the point at $(x,y)=(0,0)$ triggered by anomalous resistivity. The magnetic field shows a typical X-type topology. As a result of reconnection, the magnetic energy is converted into both thermal and kinetic energy. Two magnetic islands (so-called plasmoids) which correspond to the high-density region in Fig. 15 move in opposite directions (the figure shows only half of the simulation region and only one magnetic island) and are accelerated along the direction of the magnetic field.     
\begin{figure}[h!]
\epsscale{1.0}
\plotone{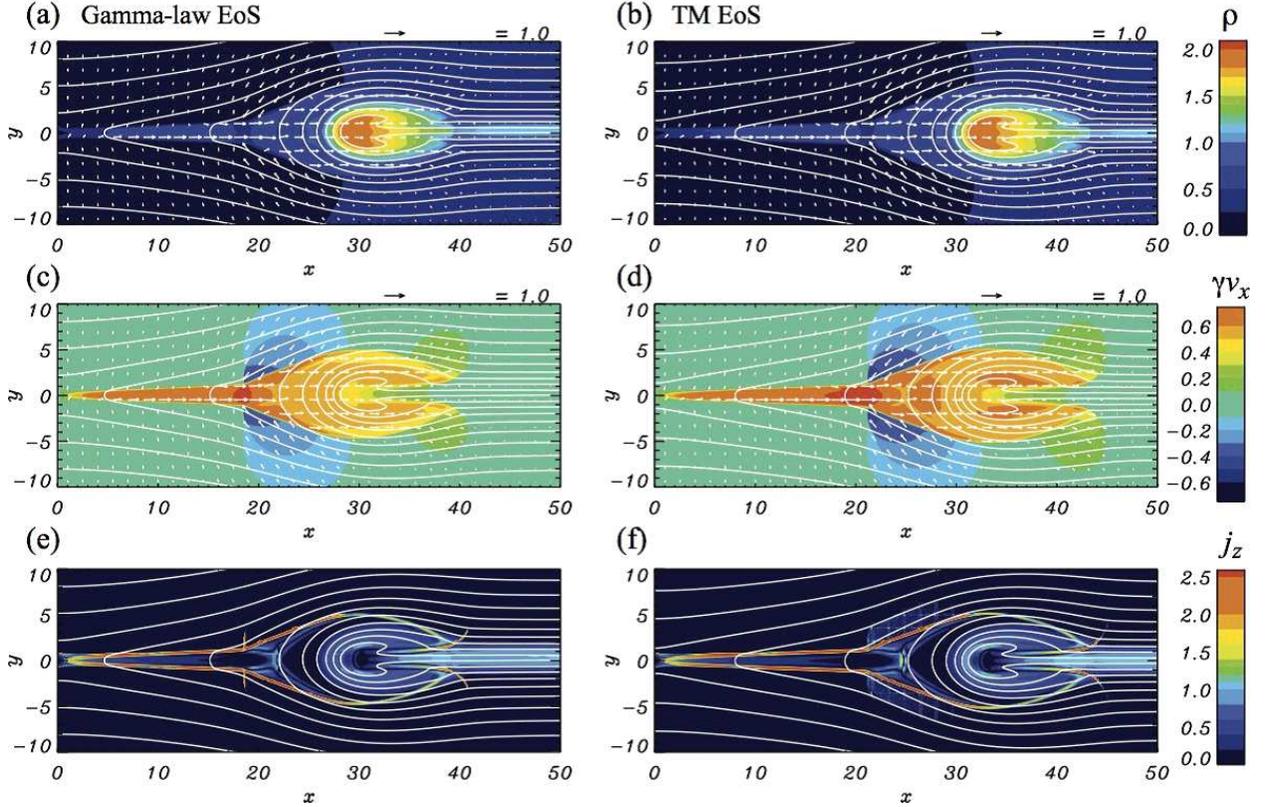}
\caption{Density (upper panels),  the $x$-component of the four-velocity $\gamma v_{x}$ (middle panels), and the out-of-plane current $j_{z}$ (lower panels) in the relativistic magnetic reconnection test at $t=100$ using the ideal EoS with $\Gamma=4/3$ (left panels) and the $TM$ EoS (right panels). White lines indicate magnetic field lines and the arrows show velocity vectors. \label{f15}}
\end{figure}
A fast reconnection jet is formed inside a narrow nozzle within a pair of slow shocks (Petschek slow shock). The reconnection jet collides with a plasmoid in front of the current sheet further downstream. The plasmoid is surrounded by strong currents (see Fig. 15e,f), which also correspond to the slow shocks (Ugai 1995). These slow shocks surround the plasmoid and are connected to the Petschek slow shocks. In the TM EoS cases, the plasmoid has a faster speed and than in the ideal EoS case with $\Gamma=4/3$, and the plasmoid in the TM EoS case propagates further. 

The time evolution of the maximum outflow velocity ($v_{x, max}$), the volume-averaged magnetic energy ($B^{2}$) and the normalized reconnection rate is shown in Figure 16.
\begin{figure}[h!]
\epsscale{0.7}
\plotone{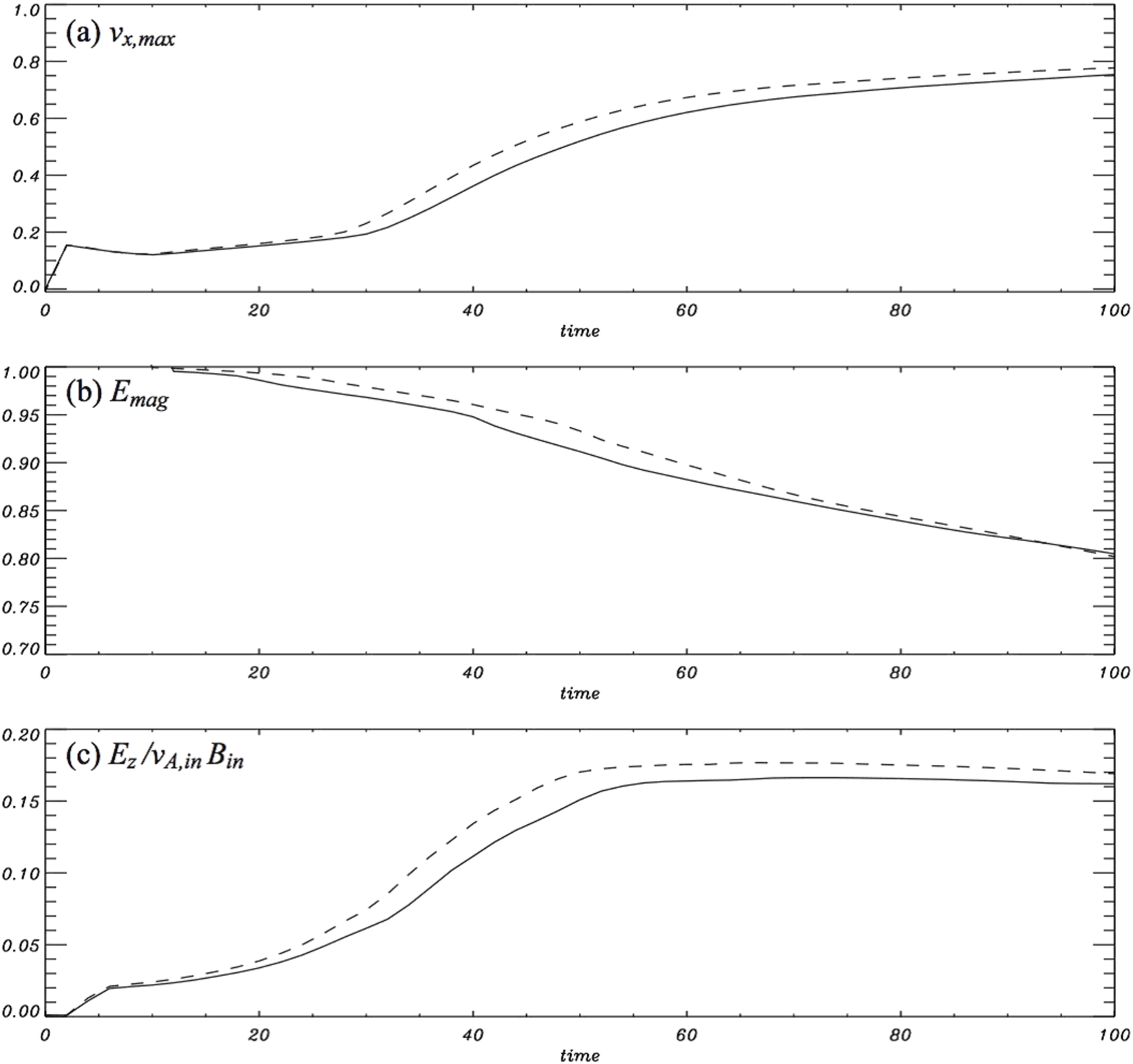}
\caption{ Evolution of (a) the maximum of the $x$-component of the velocity ($v_{x}$), (b) the volume-averaged magnetic energy ($B^{2}$), and (c) the reconnection rate as a function of time for the 2D relativistic magnetic reconnection test using an ideal EoS with $\Gamma=4/3$ (solid lines) and a TM EoS (dashed lines). \label{f16}}
\end{figure}
The outflow gradually accelerates and nearly saturates by $t \sim 60$ with $v_{x} \sim 0.8c$. The figure clearly shows that the outflow speed in the TM EoS case is slightly faster than in the ideal EoS case. The magnetic energy gradually decreases with time. Dissipated magnetic energy results in an increase to both the thermal and kinetic energy. Increase in the kinetic energy accompanies acceleration of the outflow (plasmoid). The normalized reconnection rate is defined as $\mathcal{R}=E_{z}^{*}/v_{A, in} B_{in} \sim v_{in}/v_{out}$, where $E_{z}^{*}$ is the electric field at the reconnection point and the upstream properties with subscript $in$ are evaluated at $(x,y)=(0,3)$ (Zenitani et al. 2010)\footnote{Zanotti \& Dumser (2011) and Takahashi et al. (2011) have used a different definition for late reconnection.}. The reconnection rate saturates at about $t = 50$ in both cases. However, the TM EoS case has a larger reconnection rate than the ideal EoS case ($\mathcal{R} \sim 0.16$ in the ideal EoS case with $\Gamma=4/3$ and $\mathcal{R} \sim 0.17$ in the TM EoS case). Therefore the different EoSs lead to a quantitative difference in relativistic magnetic reconnection.

\section{Summary and Conclusions}

The role of the EoS in resistive relativistic MHD using a newly developed resistive relativistic MHD code has been investigated. A number of numerical tests in 1D and multi-dimensions have been performed to check the robustness and accuracy of the new code. All of the tests show the effectiveness of the new code in situations involving both small and large uniform conductivities. 

The 1D tests of the propagation of a large amplitude circularly-polarized Alfv\'{e}n wave show the new resistive relativistic MHD code reproduces ideal relativistic MHD solutions when the conductivity $\sigma$ is high and that the code has $2^{nd}$ order accuracy. The 1D self-similar current sheet tests indicate that the analytical solution in the moderate conductivity regime is well described by the new code. In a simple MHD version of the Brio and Wu 1D shock-tube test, the code is stable using Strang's splitting technique even when the conductivity is high ($\sigma \sim 10^{6}$). In the 2D cylindrical explosion tests, at the high conductivity $\sigma \ge 10^{3}$, the results recover the solution from ideal RMHD. The results of the Kelvin-Helmholtz instability test show that the growth rate of KHI is independent of the conductivity, except for very low conductivity ($\sigma \simeq 0$). 
However, magnetic field amplification via stretching of the main vortex developed by KHI strongly depends on the conductivity.  The effect of conductivity on magnetic field amplification via KHI in 3D is a topic for future study.  

EoSs proposed by Mignone et al. (2005) and by Ryu et al. (2006), which closely approximate Synge's single-component perfect relativistic gas EoS, have been incorporated in the code. In the limit of non-relativistic and ultra-relativistic temperatures, the equivalent specific heat ratio associated with the EoSs that approximate Synge's EoS appropriately changes from the $5/3$ to the $4/3$ limits. 

The numerical tests studied the effect of the EoS on shocks, blast waves, the Kelvin-Helmholtz instability, and relativistic magnetic reconnection. The results provide a useful guide for future more specific studies of each topic. The tests confirm the general result that large temperature gradients cannot be properly described by an ideal EoS with a constant specific heat ratio. The results using a more realistic EoS, which we have studied here, show considerable dynamical differences. The 1D shock tube tests (Balsara Test2) show that the results obtained from the TM EoS and RC EoS cases are considerably different from the constant $\gamma$-law EoS case with $\gamma = 5/3$. In the 2D Kelvin-Helmholtz instability tests, the non-linear behavior depended on the EoS, even though the growth rate of the KHI was almost the same. In reconnection tests, the approximate EoS cases resulted in a faster reconnection outflow speed and a larger reconnection rate than the ideal EoS case.  We conclude that any studies of shocks, instabilities, and relativistic magnetic reconnection should use a realistic approximation to Synge's EoS.  

\acknowledgments

Y.M. would like to thank H. Takahashi, S. Zenitani, K.-I. Nishikawa and P. E. Hardee for useful comments and discussion. This work has been supported by NSF awards AST-0908010 and AST-0908040, NASA award NNX09AD16G and the Taiwan National Science Council under the grant NSC 100-2112-M-007-022-MY3. Y.M. also acknowledges partial support from the NAOJ Visiting Scholar Program (Short-term). The simulations were performed on the Columbia Supercomputer at the NAS Division of the NASA Ames Research Center, the SR16000 at YITP in Kyoto University, and Nautilus and Kraken at the National Institute for Computational Sciences in the XSEDE project supported by National Science Foundation.

\end{document}